\documentclass{article}

\usepackage{amssymb,amsfonts,amsmath,stmaryrd}
\usepackage{cite,enumerate,float}
\usepackage{color}
\def\be{\begin{eqnarray}}
\def\ee{\end{eqnarray}}

\def\tr{{\rm tr}\,}

\def\s{\theta}

\definecolor{red}{rgb}{1,0,0}
\definecolor{orange}{rgb}{1,0.5,0}
\definecolor{violet}{rgb}{0.7,0,1}

\hoffset= - 1.0in         % switch off for draft style
\newcommand\lla{\left\langle}
\newcommand\rra{\right\rangle}

\newcommand\lcb{\left\{}
\newcommand\rcb{\right\}}
\newcommand\cbr[1]{ \lcb #1 \rcb }
\newcommand\abr[1]{ \lla #1 \rra }

\newcommand\br[1]{
  \left(#1\right)
}

\newcommand\sbbr[1]{
  \left\llbracket #1 \right\rrbracket
}
\newcommand\ux[0]{\underline{x}}

\newcount\hshift
\newcount\vshift
\newcount\tmp
\newcount\i
\newcount\j
\newcount\boxsize
\newcount\circleRadius
\newcount\xShift
\newcount\yShift
\newcount\len
\newcount\tmp
\newcount\prevElt
\newcount\curElt

\def\gridPut(#1,#2)#3{{
    \loccount\x
    \x=#1
    \multiply\x by \boxsize
    \loccount\y
    \y=#2
    \multiply\y by \boxsize
    \put(\x,\y){
      #3
    }
}}

\makeatletter
\newcommand\YD[1]{{
    \vshift=0
    \@for \elt:=#1 \do{
      \Yrow\vshift\elt
      \advance\vshift by -\boxsize
    }
}}
\makeatother

\def\Ybox{{
    \let\s\boxsize
    \put(\xShift, \yShift) {
      \put(0,0){\line(1,0){\s}}
      \put(\s,0){\line(0,1){\s}}
      \put(\s,\s){\line(-1,0){\s}}
      \put(0,\s){\line(0,-1){\s}}
    }
}}

\newcommand\Yrow[2]{{
    \hshift = 0
    \j = 0
    \loop \ifnum\j<#2
    \put(\hshift,#1){\Ybox}
    \advance\hshift by \boxsize
    \advance\j by 1
    \repeat
}}

\newcommand\Ycolumn[2]{{
    \vshift = 0
    \j = 0
    \loop \ifnum\j<#2
    \put(#1,\vshift){\Ybox}
    \advance\vshift by \boxsize
    \advance\j by 1
    \repeat
}}

\newcommand\youngEnv[1]{{
    \boxsize=10
    \circleRadius=5
    \xShift=0
    \yShift=0
    #1
}}

\makeatletter
\newcommand\inlineYD[1]{{
    \yShift=0
    \xShift=0
    \def\onlyFirst{\xShift=\elt \def\onlyFirst{}}
    \@for \elt:=#1 \do{
      \onlyFirst
      \advance\yShift by 1
    }
    \multiply\yShift by 5
    \multiply\xShift by 5
    \begin{picture}(\xShift,\yShift)(0,-\yShift)
      \youngEnv{\boxsize=5 \yShift=-5 \YD{#1}}
    \end{picture}
}}
\makeatother

\makeatletter
\newsavebox{\@brx}
\newcommand{\llaa}[1][]{\savebox{\@brx}{\(\m@th{#1\langle}\)}%
  \mathopen{\copy\@brx\kern-0.5\wd\@brx\usebox{\@brx}}}
\newcommand{\rraa}[1][]{\savebox{\@brx}{\(\m@th{#1\rangle}\)}%
  \mathclose{\copy\@brx\kern-0.5\wd\@brx\usebox{\@brx}}}
\makeatother

\newcommand\aabr[1]{ \llaa #1 \rraa }

\begin{document}

\title{\vspace{1.5cm}\bf
  Towards mixed phase correlators in monomial matrix models.
}

\author{
A. Popolitov$^{a,b,c}$\footnote{popolit@gmail.com}
}

\date{ }

\maketitle

\vspace{-6cm}

\begin{center}
  \hfill MIPT/TH-24/25\\
  \hfill ITEP/TH-29/25\\
  \hfill IITP/TH-32/25
\end{center}

\vspace{4.5cm}

\begin{center}
$^a$ {\small {\it MIPT, Dolgoprudny, 141701, Russia}}\\
$^b$ {\small {\it NRC ``Kurchatov Institute", 123182, Moscow, Russia}}\\
$^c$ {\small {\it Institute for Information Transmission Problems, Moscow 127994, Russia}}
\end{center}

\vspace{.1cm}

\begin{abstract}
  Correlators in monomial Hermitian matrix model strongly depend on the choice of
  eigenvalue integration contours. We express Schur correlator in case of several different
  integration contours (mixed phase case) as a sum over products of Schur correlators
  for just one type of contour (pure phase), where expansion coefficients are
  manifestly made from Littlewood-Richardson and Mugnaghan-Nakayama coefficients.
  \\
  Further, for pure phase Schur correlators we find concise superintegrability formulas
  that unify both usual and exotic cases, that before looked very different from one another.
\end{abstract}

\bigskip

\newcommand\smallpar[1]{
  \noindent $\bullet$ \textbf{#1}
}

\section{Introduction} \label{sec:introduction}

We continue the program of studying emergent structures in strongly coupled matrix models,
as they should provide valuable insight for the emergent structures of more general QFTs
\cite{paper:BK-mm-1,paper:GM-mm-2}.
To this end, we consider the simplest essentially non-Gaussian
(i.e. interacting) matrix model -- the monomial Hermitian matrix model (MHMM):
it is a 1-matrix model with potential equal to just a monomial
$$V(X) = \tr X^r$$.

It is well-known that the measure alone does not fully specify averages in such model,
and arising extra ambiguity can be fixed, for instance, by specifying
integration contours for each of the matrix' eigenvalues, where contours should start and
end in the Stokes' sectors of the potential $V(X)$.

The space of proper contours is finite-dimensional, and for the MHMM concrete basis
of star-shaped contours $C_{r,a},\ a=1..r-1$
was proposed in \cite{paper:CHPS}. The advantage of this
set of contours is that one-dimensional moments acquire particularly simple form
(see Section~\ref{sec:definitions}). Moreover, as a result, in the case when all eigenvalues
are integrated over one and the same contour $C_{r,a}$ -- the so-called pure phase case,
the model exhibits superintegrable properties \cite{paper:MM-superint-summary}: its
partition function (the non-normalized average of $1$) is fully factorized and manifestly
expressed as a product of $\Gamma$-functions; and its (normalized) averages of Schur functions
are expressed as peculiar products of the tell-tale quantity $N-i+j$ over certain diagonals
of Young diagram $R$. Furthermore, this superintegrability of MHMM, in \cite{paper:CHPS},
was derived and explained from straightforward ``orbifoldization'' construction --
which, in principle, opened a possibility to derive similar formulas for more complicated
(e.g. polynomial) potentials.

A number of notable developments occured in understanding of MHMMs since the initial paper
\cite{paper:CHPS} sparked interest to the topic. Firstly, the case of pure phase with
non-trivial $r$-core (see App.A of \cite{paper:CHPS} for a reminder on $r$-cores
and $r$-quotients), the so-called \textit{exotic phase} was analyzed in
\cite{paper:BP-exotic}: it was discovered that, for every choice of contour $a$ and remainder
of $N \text{ mod } r =: b$ only partitions with concrete ($a,b$-dependent) rectangular
$r$-core have non-zero (non-normalized) averages. The average of Schur polynomial of the said
core then naturally plays a role of normalization constant (i.e. the partition function).

Secondly, in \cite{paper:CMPT-mon-bilinear} it was discovered that strict superintegrability
-- the property that certain bilinear averages of Schur functions $S_R$ with
auxiliary $K_\Delta$ polynomials have explicit expressions via skew Schur functions
$S_R/\Delta$, can be generalized to the case of MHMM from the Gaussian ($r=2$) model case
\cite{paper:MM-bilinear-superint}, with some reservations.

Thirdly, in \cite{paper:MM-uglov} the $\beta$-deformation of MHMM was studied. It was
discovered that this deformation is actually an Uglov matrix model
\cite{paper:U-yangian-gz}, and superintegrability
is then naturally expressed in terms of Uglov polynomials.
Mishnyakov and Myakutin considered only the case of trivial $r$-cores, but crucially
they managed to express the peculiar product over Young diagram diagonals as
ratio of Uglov polynomials evaluated at certain special locus
-- the long-awaited simplification
of the original ``orbifold'' formula in \cite{paper:CHPS}, which expressed this product
in terms of $r$-quotients $R^{(i)}, i=0..r-1$ of the diagram $R$.

\bigskip

In this paper we consider the case of mixed phase of MHMM -- that is, the case when
integration contours for different eigenvalues are different.
We try to do it honestly at \textit{finite} values of group multiplicites $N_i$,
in contrast with standard Dijkgraaf-Vafa phase approaches \cite{paper:MMSh-civ-dv},
where the perturbative expansion in $\frac{1}{N_i}$ at $N_i \rightarrow \infty$ is of main
interest.
Immediately we discover that this case is much more complicated than its pure phase counterpart: simple factorization
of Schur averages into $N$-linear factors, characteristic of pure phase superintegrable
formlulas\cite{paper:MM-superint-summary}, is lost. Still, complications remain relatively
localized: the only new complicated quantity are the expansion coefficients of
the vandermonde interaction terms (see Section~\ref{sec:vdm-int}) into Schur functions
\eqref{eq:vdm-exp-coeffs}. Still, we manage to find an expression for these expansion coefficients in terms of symmetric group characters and peculiar $N_i$-dependent conjugation
operation \eqref{eq:crq-expansion-final},\eqref{eq:prime-operation}. Our analysis naturally leads to definition of
\textit{normalized} mixed phase correlator as a sum over products of normalized pure phase
correlators\eqref{eq:main-formula-1}.

For these latter we manage to improve upon result of Mishnyakov-Myakutin, and express
pure phase Schur correlator in \textit{both} usual and exotic phases (that is, when partition
has non-trivial $r$-core and remainder of $N \mod r$ is adjusted accordingly) as (ratio of)
skew Schur functions at certain special locus \eqref{eq:pure-phase-final}
-- bringing MHMM close (in form) o the celebrated WLZZ family of superintegrable
matrix models.

The formulas \eqref{eq:main-formula-1} and \eqref{eq:pure-phase-final} are
\textbf{the main results} of the present paper. After necessary reminder of definitions
and notation (Section~\ref{sec:definitions}), the former (Section~\ref{sec:vdm-int})
and the latter (Section~\ref{sec:box-product-formula}) are presented.

%% All these previous developments restricted themselves to the pure phase case.
%% In this paper we do the next logical step and consider the case when several distinct integration contours are chosen -- starting from the case of just two distinct (groups of) contours.
%% After a necessary reminder of definitions (Section~\ref{sec:definitions})
%% we demonstrate (???)

\section{Definitions and notations} \label{sec:definitions}

\begin{itemize}
\item A collection of numbers is often denoted with underline, for instance
  \begin{align}
    \underline{x} := \br{x_1,\dots,x_N} \ \ \ \ 
    \underline{x}^{(m)} := \br{x^{(m)}_1, \dots, x^{(m)}_{N_m}},
    \ \ \ \ \underline{a} := \br{a_1,\dots,a_m}
  \end{align}
  Note how the cardinality of the collection is assumed from the context.
\item By Schur polynomial with argument in curly braces we mean
  Schur polynomial in time variables, where curly brace argument depends on index $k$
  and is the value of time $p_k$, in particular
  \begin{align}
    S_R\cbr{\sum_i x_i^k} \equiv S_R\br{\underline{x}}
  \end{align}
\end{itemize}

\bigskip

\noindent Following \cite{paper:CHPS}, non-normalized averages in monomial Hermitian matrix model (MHMM) are defined as follows
\begin{align} \label{eq:mm-int}
  \lla f(x_1,\dots x_N) \rra :=
  \int_{\mathcal{C}_1} \dots \int_{\mathcal{C}_N}
  d x_1 \dots d x_N \underbrace{\prod_{i<j}(x_i-x_j)^2}_{\Delta^2}
  \exp\br{-\sum_{i=1}^N x_i^r} f(x_1,\dots x_N)
\end{align}
There are $r$ Stokes sectors (directions at infinity) for this integral and, hence,
the space of possible integration contours $\mathcal{C}_i$ is $r-1$-dimensional. One smart choice of
basis in this space is provided by the star-shaped contours
\begin{align}
  C_{a} = \sum_{j=0}^{r-1} \omega^{-a j} \cdot [0,\ \omega^j \infty), a = 1 .. r-1,
\end{align}
where $\omega = \exp\br{\frac{2 \pi i}{r}}$ is the $r$-th primitive root of unity
and $\omega^j \infty$ means the contour goes to infinity in the direction of $\omega^j$.

The advantage of this basis is that one-dimensional moments are very simple
\begin{align} \label{eq:1d-moment-def}
  \mathbb{M}_{a}(p) := \int_{C_a} d x x^p \exp\br{-x^p} = \delta_{r|p+1-a}
  \Gamma\br{\frac{p+1}{a}},
\end{align}
where $\delta_{r|p+1-a}$ equals $1$ whenever $p+1-a = 0 \mod r$ and $0$ otherwise.
At this point (and in what follows) it is also quite natural to include into consideration
the case $a = 0$ as well: the r.h.s. of \eqref{eq:1d-moment-def} does not single out this
case in any way and, as we will see further, all the relevant
superintegrability structures also seamlessly interpolate to the $a = 0$ case.\footnote{
Let us stress that the contour $C_0$ is not a closed contour -- the apparent boundary point
$0$, which cancels for all other values of $a \neq 0$, is present for $a=0$.
Hence, the moments \eqref{eq:1d-moment-def}, and all the multidimensional $N>1$ moments that
can be defined from them by induction, are not solutions to (naive) Ward identities
for the MHMM \eqref{eq:mm-int}. However, the explanation of why superintegrability persists
in this case is the following. One can include a Fayet-Illiopoulos (FI) term
$\prod_{i=1}^N x_i^u$ into integrand, and the point $x = 0$ is then one of the zeroes
of the measure for positive integer $u$ -- and hence in this more general setting the contour $C_0$ is the allowed one; and the Ward identities are modified accordingly.
From this point of view our formulas for $C_0$ correspond to putting $u = 0$ in the analytically continued answers from positive integer $u$, which is straightforward to verify
in concrete examples but may be tedious to prove in general.}

\bigskip

The simplest case of contour choice is when all contours $\mathcal{C}_i$ are equal to
one and the same $C_a$ -- the so-called \textbf{pure phase} case.
%% \texttt{Definition of N > 1 correlators in Schur basis.}
Averages of Schur polynomials $S_R(x_1,\dots, x_N)$ \cite{book:M-sym-fun}
in pure phase have a number of striking properties:

\begin{enumerate}
\item Schur averages depend \textit{quite strongly} on the remainder $b$ of $N$ modulo
  $r$
  $$ N = r k + b, k \in \mathbb{Z} $$
\item for the most straightforward cases of $b = 0 \text{ or } a$ this normalized
  average is, in fact, given by nice concise formula \cite{paper:CHPS}
\begin{align}\label{eq:mm-superint}
  \llaa S_R(x_1,\dots, x_N) \rraa_a := \frac{\lla S_R(x_1,\dots, x_N) \rra_a}{\lla 1 \rra_a}
  = S_R\br{p_k=\delta_{k,r}} \prod_{(i,j)\in R}
  \br{\llbracket N; - i + j\rrbracket_{r,0}
  \cdot \left\{
  \begin{array}{l}
    \llbracket N; - i + j\rrbracket_{r,a} \text{ if } b = 0 \\
    \llbracket N; - i + j\rrbracket_{r,r-a} \text{ if } b = a \\
  \end{array}
  \right . },
\end{align}
where $\llbracket N; - i + j\rrbracket_{r,a}$ is the peculiar selection factor
that is nontrivial only for certain diagonals of the Young diagram $R$.
\footnote{
Note how if one puts in the formulas $a = 0$ the two cases for different $b$
correctly coincide with one another
\begin{align}
  \llaa S_R(x_1,\dots, x_N) \rraa_0
  = S_R\br{p_k=\delta_{k,r}} \prod_{(i,j)\in R}
  \br{\llbracket N; - i + j\rrbracket_{r,0}}^2
\end{align}
}
\begin{align}
  \llbracket N; w \rrbracket_{r,A} :=
  \left\{
  \begin{array}{l}
    (N + w) \text{ if } w = A \mod r \\
    1 \text{ otherwise}
  \end{array}
  \right .
\end{align}

\item For more complicated case of generic $b \neq 0,a$ the average of unity
  $\abr{1}_a$ is zero, and instead non-zero are the averages $\abr{S_R}$ only for
  such partitions $R$, whose $r$-core $\rho(R)$ (see App.A of \cite{paper:CHPS})
  is rectangular partition of shape $b \times (a - b)$ or $(b-a) \times (r-b)$, depending on whether $b < a$ or $b > a$
  \cite{paper:BP-exotic}. Then it makes sense to define normalized averages by
  \begin{align}
    \aabr{S_R}_a := \abr{S_R}_a / \abr{S_{\rho(R)}}_a
  \end{align}
\end{enumerate}
which generalizes \eqref{eq:mm-superint}, since $\rho(R) = \emptyset$
(and hence $S_{\rho(R)} = 1$) for such partitions that selection factor
$S_R\cbr{\delta_{k,r}}$ is nonzero.

The main result of \cite{paper:BP-exotic} (see Eqn.(21)) expresses the
normalized average $\aabr{S_R}_a$ through Schur polynomials of $r$-quotients $R^{(i)}$.
\begin{align}\label{eq:schur-avg-exotic-explicit}
  \aabr{S_R}_a = & \
  \frac{1}{r^{(|\mu| - |\rho|)/r}
    S_{R/\rho(R)}\cbr{\delta_{k,r}}}
  \prod_{j=0}^{r-1}
  S_{R^{\br{m_r(j + m_r(b-a))}}}
  \cbr{\frac{1}{r}\left( N
  + \delta_{j < b}\cdot m_r(-b) + \delta_{j \geq b}\cdot n_r(-b) \right)
  }
  \\ \notag & \times
  \prod_{j=0}^{r-1}
  S_{R^{(m_r(j+b))}}
  \cbr{\frac{1}{r}\left( N +
  \delta_{j < m_r(b-a)}\cdot n_r(a-b) + \delta_{j \geq m_r(b-a)}\cdot m_r(a-b)
  \right)}
\end{align}
in somewhat involved way using peculiar remainder functions
  \begin{align}
    m_r(x) = & \ x \mod r \in [0, r-1] \text{ and } \\ \notag
    n_r(x) = & \ m_r(x) - r \in [-r, -1]
  \end{align}
  in this paper we obtain much more clear \eqref{eq:pure-phase-final} that expresses
  everything through original diagram $R$ and its core $\rho(R)$, without the need to
  explicitly involve quotients $R^{(i)}$.

\bigskip

In the more complicated \textbf{mixed phase case} the eigenvalues $x_1,\dots x_N$ are grouped
into $m \geq 2$ groups of sizes $N_1,\dots N_m$, respectively, each of which
is integrated along the respective contour $C_{a_1}, \dots C_{a_m}$.

Since one can now distinguish between the eigenvalues belonging to different groups,
but still cannot distinguish the ones in the same group, it makes sense that
now we are interested in (non-normalized) multi-Schur correlators
\begin{align} \label{eq:mixed-phase-corr}
  \lla
  \prod_{j=1}^m S_{R_j}\br{\ux^{(j)}}
  \rra_{a_1,\dots,a_m}
\end{align}
where each Schur polynomial depends on eigenvalues from its own group $j$.
It is a separate question (discussed in Section~\ref{sec:vdm-int})
of how to define the analogs of the normalized correlators in a meaningful way.
In what follows we frequently omit the contour index $a_1,\dots a_m$, spelling it out
only when necessary.

\section{Van-der-monde interaction term} \label{sec:vdm-int}

%% ### vv ??? Once we established that the structure is a moot one -- we can proceed to building more realistic constructions ???

In the mixed phase setting the $N$ eigenvalues are split into $m$ groups
with multiplicities $N_1,\dots ,N_m$
\begin{align}
  \br{x_1,\dots,x_N} = \br{\underbrace{x^{(1)}_1,\dots,x^{(1)}_{N_1}}_{\text{group 1}},
  \dots
  \underbrace{x^{(m)}_1,\dots,x^{(m)}_{N_m}}_{\text{group $m$}}}
\end{align}

Correspondingly, the (square of) vandermonde determinant $\Delta^2$
in \eqref{eq:mm-int} splits into product of vandermonde determinants for the groups
and pairwise interaction terms between the groups
\begin{align}\label{eq:delta-2-groups}
  \Delta^2 = \prod_{i<j}(x_i-x_j)^2
  = \prod_{l=1}^m \underbrace{\prod_{i<j}^{N_l}
    \br{x^{(l)}_i-x^{(l)}_j}^2}_{\br{\Delta^{(l)}}^2}
  \cdot
  \prod_{l_1 < l_2}^m \underbrace{\prod_{i=1}^{N_{l_1}}
    \prod_{j=1}^{N_{l_2}}\br{x^{(l_1)}_i-x^{(l_2)}_j}^2}_{\br{\Delta^{(l_1,l_2)_{\text{int}}}}^2}
\end{align}

%% Our goal in this section is to describe, as concretely as possible, decomposition
%% for these interaction terms into products of Schur functions. As to not overload presentation
%% with superscripts, in this section we consider two groups with multiplicities $N$ and $M$;
%% eigenvalues of the first group are denoted as $\underline{x}$ and the second group
%% -- as $\underline{y}$. We are after description of coefficient $c_{R,Q}$ in the decomposition

Pairwise interaction terms in \eqref{eq:delta-2-groups} are symmetric functions in variables
belonging to each of the two interacting groups, so they should be expandable in the
Schur basis
\begin{align}\label{eq:vdm-exp-coeffs}
  \prod_{i=1}^N \prod_{j=1}^M \br{x_i - y_j}^2
  = \sum_{R,Q} \ c_{R,Q} \ S_R\br{\underline{x}}S_Q\br{\underline{y}}
\end{align}
The expansion coefficients $c_{R,Q}$ are then crucial ingredient for
reducing the mixed-phase Schur correlators to the (sum of) products of pure phase Schur
correlators -- for which we do know explicit formulas.
Namely, observe that in the mixed phase correlator \eqref{eq:mixed-phase-corr}
\textit{the only} interaction between eigenvalues' groups is given by the vandermonde
interaction terms. Hence, one may write, in case of just two groups
(the case of $m$ groups is analogous)
\begin{align} \label{eq:multi-schur-explicit}
  \lla
  S_{R^{(1)}}\br{\underline{x}^{(1)}}
  S_{R^{(2)}}\br{\underline{x}^{(2)}}
  \rra_{a_1,a_2}
  = \sum_{P^{(1)},P^{(2)}}
  \sum_{Q^{(1)},Q^{(2)}}
  c_{P^{(1)},P^{(2)}}
  N_{R^{(1)}P^{(1)}}^{Q^{(1)}}
  N_{R^{(2)}P^{(2)}}^{Q^{(2)}}
  \lla
  S_{Q^{(1)}}\br{\underline{x}^{(1)}}
  \rra_{a_1}
  \lla
  S_{Q^{(2)}}\br{\underline{x}^{(2)}}
  \rra_{a_2},
\end{align}
where $N_{R P}^{Q}$ are the Littlewood-Richardson coefficients
\begin{align}
  S_R(\underline{x}) S_P(\underline{x}) = \sum_Q N_{R P}^{Q} S_Q(\underline{x})
\end{align}
An analogous formula is straightforward to write down in the case of arbitrary $m$
number of groups.

\bigskip

We, therefore, need as explicit a formula for the coefficients $c_{R,Q}$, as possible,
which we now describe.
First, let us observe that \eqref{eq:vdm-exp-coeffs} is similar to Cauchy formula,
which states that \cite{book:M-sym-fun}
\begin{align} \label{eq:cauchy-formula}
  \prod_{i=1}^N \prod_{j=1}^M \br{1 - x_i y_j}^{-1} = \sum_{R,Q}
  \delta_{R,Q} S_R\br{\underline{x}} S_Q\br{\underline{y}},
\end{align}
where automatically the sum is truncated to only those Young diagrams $R = Q$
for which $l(R) \leq N$
and $l(R) \leq M$, $l(R)$ being the length of the diagram.

Secondly, from \eqref{eq:cauchy-formula} it readily follows that there is
a simple formula for $\Delta_{\text{int}}$ (i.e. the non-squared interaction term)
\begin{align} \label{eq:delta-int}
  \prod_{i=1}^N \prod_{j=1}^M \br{x_i - y_j} = \sum_{R \sqcup Q^T = \text{Rectangle}(N,M)}
  (-1)^{|Q|} S_R\br{\underline{x}} S_Q\br{\underline{y}},
\end{align}
that is the sum goes over such pairs of Young diagrams $R$ and $Q$ that
$R$ and the \textit{transpose} of $Q$ form an $N$ by $M$ rectangle.

For the $\Delta_{\text{int}}^2$, which actually enters \eqref{eq:vdm-exp-coeffs}
one analogously gets
\begin{align} \label{eq:delta-int-squared}
  \prod_{i=1}^N \prod_{j=1}^M \br{x_i - y_j}^2
  = & \ \prod_{i=1}^N x_i^{2 M} \prod_{j=1}^M \br{1 - \frac{y_j}{x_i}}^2
  \mathop{=}_{\substack{\text{Cauchy}\\\text{formula}}}
  \prod_{i=1}^N x_i^{2 M} \exp\br{-2 \sum_k
    \frac{p_k(\underline{y}) p_k(\underline{x})}{k}
  } \\ \notag
  = & \   \br{\prod_{i=1}^N x_i^{2 M}}
  \sum_R S_R\cbr{-2\sum_i y_i^k}
  S_R\cbr{\sum_i \frac{1}{x_i^k}}
\end{align}

Since, on the one hand
\begin{align}
  S_R\left\{-p_k\right\} = S_{R^T} \left\{p_k\right\}
\end{align}
and on the other hand
\begin{align}
  \br{\prod_{i=1}^N x_i^{2 M}}
  S_R\left\{\sum_i \frac{1}{x_i^k}\right\}
  = S_{R'}\left\{\sum_i x_i^k \right\}
  , \text{ where } R' \sqcup R = \text{Rectangle}(N, 2M),
\end{align}
that is $R'$ is ``conjugate'' Young diagram to $R$ in the rectangle $N$ by $2 M$,
we can continue
\begin{align}
  \eqref{eq:delta-int-squared} =
  \sum_R S_{R^T} \cbr{2 \sum_i y_i^k} S_{R'}\br{\underline{x}}
\end{align}

The last component is an expression of $S_{R^T} \cbr{2 \sum_i y_i^k}$
through $S_Q\br{\underline{y}}$ which can be obtained from orthogonality of Schur polynomials
\begin{align}
  S_{R^T} \cbr{2 \sum_i y_i^k} =
  \sum_{\Delta, Q} \Psi_{R^T} (\Delta) \frac{2^{l(\Delta)}}{z_\Delta}
  \Psi_Q(\Delta) \cdot S_Q \br{\underline{y}},
\end{align}
where $\Psi_Q(\Delta)$ are symmetric group characters that can be calculated
with help of Murnaghan-Nakayama rule \cite{book:S-enumerative-combinatorics}
and $z_\Delta$ is the standard combinatorial normalization coefficient
in the theory of symmetric functions.

Assembling everything together, we have
\begin{align} \label{eq:crq-expansion-final}
  \prod_{i=1}^N \prod_{j=1}^M \br{x_i - y_j}^2
  = \sum_{R,Q} \ \boxed{c_{R,Q}} \ S_R\br{\underline{x}}S_Q\br{\underline{y}}
  = \sum_{R,Q} \ \boxed{\br{\sum_\Delta
    \Psi_{R^{'T}} (\Delta) \frac{2^{l(\Delta)}}{z_\Delta} \Psi_Q(\Delta)
    }} \ S_R\br{\underline{x}}S_Q\br{\underline{y}}
\end{align}
where the prime ($'$) operation is rather tricky since it implicitly depends on $N$ and $M$
\begin{align} \label{eq:prime-operation}
  R^{'} \ : \ R^{'} \sqcup R = \text{Rectangle(N,M)}
\end{align}
\noindent and does so in a non-polynomial way.

Alternatively, one may take the square of formula \eqref{eq:delta-int}
-- and then $c_{R,P}$ turns out to be expressed as a peculiar sum of
Littewood-Richardson coefficients.

\bigskip

As we see, the expression for $c_{R,P}$ turns out to be quite involved. It either  contains, as its integral component, Littlewood-Richardson or Murnaghan-Nakayama rule,
neither of which is given by
a closed formula, but rather by a combinatorial algorithm.
Also, the prime (') operation of taking the conjugate partition essentially and
strongly depends on multiplicities $N$ and $M$, and it is not obvious at all
how to express this dependence as some sort of polynomial (or even power-law) dependence.
\footnote{
Perhaps, the point-of-view on Schur polynomials through Noumi-Shiraishi functions,
where partition parts $R_i$ become arbitrary complex variables can be
fruitful here, however, this is a separate line of development that needs to be pursued
elsewhere.}
These are not drawbacks at all if one is after calculating any concrete mixed phase correlator
(that is for fixed $R^{(i)}$ and $N_i$).
However, dependence on $N$ is much more involved in this formula, as compared to
typical pure phase answer \eqref{eq:mm-superint},
so proper language needs to be designed in the future to express such dependiencies conveniently, possibly using something along the lines of
generalized Borel transform trick \cite{paper:MMPS-miwa,paper:MMPS-superint-point}.

\bigskip

Since there is no \textit{a priori} reason to expect that the non-normalized multi-Schur
correlator \eqref{eq:multi-schur-explicit} would be divisible by anything else
than the corresponding pure phase partition functions
\footnote{Where the role of partition funciton is played by either
$\lla 1 \rra_a$ or $\lla S_{c(R)} \rra_a$ -- see Section~\ref{sec:definitions}.},
it is reasonable to \textit{define} the \textbf{normalized mixed-phase correlator}
(for two groups, and, analogously -- for more groups) as
\begin{align} \label{eq:main-formula-1}
  \boxed{
    \llaa
    S_{R^{(1)}}\br{\underline{x}^{(1)}}
    S_{R^{(2)}}\br{\underline{x}^{(2)}}
    \rraa_{a_1,a_2}
    := \sum_{P^{(1)},P^{(2)}}
    \sum_{Q^{(1)},Q^{(2)}}
    c_{P^{(1)},P^{(2)}}
    N_{R^{(1)}P^{(1)}}^{Q^{(1)}}
    N_{R^{(2)}P^{(2)}}^{Q^{(2)}}
    \llaa
    S_{Q^{(1)}}\br{\underline{x}^{(1)}}
    \rraa_{a_1}
    \llaa
    S_{Q^{(2)}}\br{\underline{x}^{(2)}}
    \rraa_{a_2}
  }
\end{align}
This is in analogy with the multi-component correlators in Chern-Simons/knot theory,
where the widely accepted way to normalize Wilson loop averages over a link
is to simply divide it by the averages of the corresponding colored unknots.

\section{The box product formula} \label{sec:box-product-formula}

In \cite{paper:MM-uglov} remarkable progress was made: the formula for the peculiar
box product in \eqref{eq:mm-superint} was found that expresses it as a (ratio of)
Schur polynomials of partition $R$, evaluated at certain special locus
(c.f. \cite{paper:MM-uglov} (56))
\begin{align}\label{eq:box-product-via-schurs}
  \prod_{\substack{(i,j)\in R \\ \rho(R) = \emptyset}}
  \llbracket N; - i + j\rrbracket_{r,b}
    = \frac{S_R\cbr{\sum_{i=0}^{N-1} \omega^{i k}}}{S_R\cbr{\delta_{k,r}}}
\end{align}
Note that in this formula the argument $b$ does not appear \textit{explicitly} on the
right hand side.
Instead, it is implicit in the choice of $N$ with appropriate remainder mod $r$:
$N = r k + b$.

The formula \eqref{eq:box-product-via-schurs} is a major improvement from
the original formula for the same product
in \cite{paper:CHPS} eqn.(4.14), which expressed the same box product via Schur polynomials
of $r$-quotients $R^{(i)}$, since these $r$-quotients require additional combinatorial
calculation from $R$ via abacus diagram (see \cite{paper:CHPS}~App.~A).

Still, the formula \eqref{eq:box-product-via-schurs} can use some improvement because:
\begin{itemize}
\item Number of eigenvalues $N$ appears as upper summation index, making it difficult
  to treat $N$ as \textit{symbolic variable} (which may be desirable, for instance, in
  $W_{1+\infty}$-approach
  \cite{paper:MMMP-commutative-fams-in-winfty}
  to MMs, where $N$ is just the value of the central
  charge, and is treated ``on equal footing'' with other constants of the algebra).
\item The formula \eqref{eq:box-product-via-schurs} does \textit{not} describe
  box products needed for the exotic phase
  \cite{paper:BP-exotic} -- the denominator $S_R\cbr{\delta_{k,r}}$ vanishes
  for these cases. The currently known formula
  \eqref{eq:schur-avg-exotic-explicit}
  is still in terms of $r$-quotients $R^{(i)}$.
\end{itemize}

To improve on the first point, we, first of all, note that the sum is cyclotomic
and geometric so that in fact it can be explicitly calculated
\begin{align} \label{eq:geom-sum-taken}
  \sum_{i=0}^{N-1} \omega^{i k} \mathop{=}_{\substack{N = r l + b \\ k = r i + m}}
  N \cdot \delta_{m,0} + \br{1 - \delta_{m,0}} \frac{\omega^{m b} - 1}{\omega^m - 1},
\end{align}
where we isolated the $m=0$ case explicitly to expand the occuring
$0/0$-indeterminancy.

Secondly, \eqref{eq:geom-sum-taken} allows us to ``disentangle'' in \eqref{eq:box-product-via-schurs} the residue of $N$ mod $r$ and the shift of the diagonal, over which the
box product is taken. We have
\begin{align} \label{eq:box-product-taken}
  \boxed{
    \prod_{\substack{(i,j)\in R \\ \rho(R) = \emptyset}}
    \llbracket N; - i + j\rrbracket_{r,b}
    = \frac{S_R\cbr{N \cdot \delta_{m,0} + \br{1 - \delta_{m,0}} \frac{\omega^{m b} - 1}{\omega^m - 1}}}{S_R\cbr{\delta_{k,r}}}
    =: \frac{S_R\cbr{\pi_k^\star(m,b)}}{S_R\cbr{\delta_{k,r}}}
  }
\end{align}
where, as above $m = k \mod r$ and now $N$ is just a symbol on both sides of the equality, while the parameter $b$
on the r.h.s. directly corresponds to diagonal shift on the l.h.s.
For brevity, in what follows we denote the peculiar substitution of $p_k$ in the numerators
with $\pi_k^\star(m,b)$:
\begin{align} \label{eq:pi-locus}
  \pi_k^\star(m,b): \ \ p_k = N \cdot \delta_{m,0} + \br{1 - \delta_{m,0}} \frac{\omega^{m b} - 1}{\omega^m - 1}
\end{align}

\bigskip

Having concise and clear \eqref{eq:box-product-taken} at our hands,
we are ready to tackle the second point. The insight from \cite{paper:BP-exotic}
is that \textit{skew Schur} polynomials play instrumental role in the exotic phase,
where the ``denominator'' partition of the skew Schur polynomial is the $r$-core
of the partition. This leads us to suggest an ansatz (easily verifiable with computer experiments)
\begin{align} \label{eq:box-product-exotic-taken}
  \boxed{
  \prod_{\substack{(i,j)\in R \\ \rho(R) \neq \emptyset}}
    \llbracket N; - i + j\rrbracket_{r,b}
    = \frac{1}{S_{R/\rho(R)}\cbr{\delta_{k,r}}}
    \cdot
    \frac{S_R\cbr{\pi_k^\star(m,b)}}{
      S_{\rho(R)}\cbr{\pi_k^\star(m,b)}},
  }
\end{align}
which seems straightforward generalization of \eqref{eq:box-product-taken}.
Note, however, a thing about this formula:
\textbf{it only works where it absolutely must} for the ``needs'' of monomial MMs
-- i.e. only when
\begin{itemize}
\item the $r$-core $\rho(R)$ is rectangular partition
  (see \cite{paper:BP-exotic}~eqn.(14))
\item we are taking the product over diagonals that are immediately above and below this core partition.
\end{itemize}
These are the only products over boxes that, in fact, appear in exotic phase of MHMM
-- so formula \eqref{eq:box-product-exotic-taken} is sufficient.
Our attempts to search for further generalizations of \eqref{eq:box-product-exotic-taken}
to non-rectangular cores and/or other diagonals ``for completeness sake'' were
unsuccessful, but, again, for our current interest -- the Schur averages in MHMM,
formula \eqref{eq:box-product-exotic-taken} \textit{is} sufficient.

This situation strangely resembles currently very active developments
in Vogel universality \cite{paper:MS-vogel-review,
  paper:B-vogel-and-macdonald,
  paper:AIKM-the-uniform-structure-of-g4}
-- where universal formulas appear to be present
\textit{only} for such (combinations of) quantum and refined dimensions
that are relevant from the (refined) Chern-Simons observables point-of-view.

\bigskip

One can now assemble all the superintegrable answers for different
$a$ and $b$ (including $a=0$) into nice table. For instance, for $r=2, 3$
\begin{align}
  r = 2:
  \begin{array}{|l|l|l|}
    \hline
    a \backslash b & 0 & 1 \\ \hline
    0 & \emptyset;\ \ S_R^* \sbbr{\cdot}_0\sbbr{\cdot}_0 &
    \inlineYD{1};\ \ S_{R/\rho(R)}^* \sbbr{\cdot}_1\sbbr{\cdot}_1 \\ \hline
    1 & \emptyset;\ \ S_R^* \sbbr{\cdot}_0\sbbr{\cdot}_1 &
    \emptyset;\ \ S_R^* \sbbr{\cdot}_0\sbbr{\cdot}_1 \\ \hline
  \end{array}
\end{align}
\begin{align}
  r = 3:
  \begin{array}{|l|l|l|l|}
    \hline
    a \backslash b & 0 & 1 & 2 \\ \hline
    0 & \emptyset;\ \ S_R^* \sbbr{\cdot}_0\sbbr{\cdot}_0 &
    \inlineYD{2};\ \ S_{R/\rho(R)}^* \sbbr{\cdot}_2 \sbbr{\cdot}_2 &
    \inlineYD{1,1};\ \ S_{R/\rho(R)}^* \sbbr{\cdot}_1 \sbbr{\cdot}_1 \\ \hline
    1 & \emptyset;\ \ S_R^* \sbbr{\cdot}_0\sbbr{\cdot}_1 &
    \emptyset;\ \ S_R^* \sbbr{\cdot}_0\sbbr{\cdot}_2 &
    \inlineYD{1};\ \ S_{R/\rho(R)}^* \sbbr{\cdot}_1 \sbbr{\cdot}_2 \\ \hline
    2 & \emptyset;\ \ S_R^* \sbbr{\cdot}_0\sbbr{\cdot}_2 &
    \inlineYD{1};\ \ S_{R/\rho(R)}^* \sbbr{\cdot}_1 \sbbr{\cdot}_2 &
    \emptyset;\ \ S_R^* \sbbr{\cdot}_0\sbbr{\cdot}_1 \\ \hline
  \end{array},
\end{align}
where in each table entry first is the corresponding $r$-core (partitions with which
have non-zero averages) and second is the tersely written formula for $\aabr{S_R}$.
$S_R^*$ means $S_R\cbr{\delta_{k,r}}$ and $\sbbr{\cdot}_a$ is a shorthand for
$\prod_{(i,j)\in R} \sbbr{N-i+j}_{r,a}$

\bigskip

Furthermore, we can finally compress these answers for
different ``branches'' (ordinary and exotic) of the pure phase into \textbf{single universal formula}, thus unifying all cases
\begin{align}
  \aabr{S_R}_a = S_{R/\rho(R)}\cbr{\delta_{k,r}}
  \prod_{(i,j)\in R} \sbbr{N-i+j}_{r,\text{mod}(-b,r)}
  \sbbr{N-i+j}_{r,\text{mod}(a-b,r)}
\end{align}
that, with help of \eqref{eq:box-product-exotic-taken} is rewritten solely in terms of Schur functions at special points
\begin{align} \label{eq:pure-phase-final}
  \boxed{
    \aabr{S_R}_a =
    \frac{1}{S_{R/\rho(R)}\cbr{\delta_{k,r}}}
    \cdot
    \frac{S_R\cbr{\pi_k^\star(m,\text{mod}(-b,r))}}
         {S_{\rho(R)}\cbr{\pi_k^\star(m,\text{mod}(-b,r))}}
    \cdot
    \frac{S_R\cbr{\pi_k^\star(m,\text{mod}(a-b,r))}}
         {S_{\rho(R)}\cbr{\pi_k^\star(m,\text{mod}(a-b,r))}}
  }
\end{align}

The shape of this formula brings the ``pure phase'' case of the MHMM much closer
in spirit to the WLZZ \cite{paper:MMMPWZ-interpolating-wlzz,
  paper:MOP-beta-wlzz-directly}
models, for which the average
of Schur polynomial reads, in case the potential for matrix $Y$ (see
\cite{paper:MMMPWZ-interpolating-wlzz}(16))
begin monomial as well $V(Y) = \tr Y^r$
\begin{align} \label{eq:wlzz-avg}
  \aabr{S_R}_{\text{WLZZ}} = \sum_{Q} S_{R/Q}\cbr{\delta_{k,r}}
  \frac{S_R\cbr{N} S_Q\cbr{\delta_{k,1}}}
       {S_R\cbr{\delta_{k,1}} S_Q\cbr{N}} S_Q\cbr{g_k},
\end{align}
where $\bar{p}_k$ and $g_k$ are the generating parameters of the model.
This brings about very interesting question on
\begin{itemize}
\item
  Whether more than one superintegrable branch, essentially different from one another
  may be present for the same eigenvalue matrix model?
\item If so, what is the difference between the corresponding $W$-operator representation?
  How the two distinct representations appear from the same set of Ward identities?
\end{itemize}
In \cite{paper:MOP-w3+2} authors already made an attempt to analyze space of solutions of
(monomial) WLZZ models, and newly found \eqref{eq:pure-phase-final} makes this
research direction even more worthwhile.

\subsection{Towards character phase of Kontsevich model}

Further, similarity between \eqref{eq:wlzz-avg} and \eqref{eq:pure-phase-final}
suggests the following development direction. In \cite{paper:MM-konts-super} the
superintegrability of the celebrated Kontsevich cubic model was established,
where the superintegrability is in terms of $Q$-Schur functions
\begin{align} \label{eq:kontsevich-superint}
  Z_K \mathop{=}_{\substack{\text{Kontsevich} \\ \text{phase}}}
  \sum_{R \in \cbr{\substack{\text{Strict} \\ \text{partitions}}}}
  \frac{1}{4^{|R|}} \frac{Q_R\cbr{\delta_{k,1}}
    Q_{2 R}\cbr{\delta_{k,3}}}
       {Q_{2 R} \cbr{\delta_{k,1}}}
       \cdot Q_R\cbr{p_k}
\end{align}

Kontsevich phase, and its answer \eqref{eq:kontsevich-superint}, correspond to the regime
where Gaussian term is dominant, and cubic vertex is considered perturbatively. However, there is another regime -- the character phase, where cubic term si the main one, and in this regard Kontsevich model may be viewed as monomial matrix model; but with times introduced differently, via Miwa parametrization $t_k = \frac{1}{k} \tr \Lambda^{-k}$.
One may, therefore, wonder, whether similarly explicit analog of \eqref{eq:pure-phase-final}
exists for character phase of KM (and Generalized KM) and how it interpolates to
the Kontsevich phase answer \eqref{eq:kontsevich-superint}.

\section{Conclusion} \label{sec:conclusion}

In this paper we performed the first necessary steps towards
non-perturbative description of mixed-phase correlators in matrix models
with simplest non-Gaussian potential -- the monomial one.

Firstly, we expressed mixed phase correlator as a sum over products of pure phase
correlators (c.f. \eqref{eq:main-formula-1},\eqref{eq:crq-expansion-final}),
and we demonstrated that their explicit description involves either Littlewood-Rchardson,
or Murnaghan-Nakayama coefficients. This, on one hand, means that discovering
a formula for mixed-phase correlators in completely elementary terms is highly unlikely,
but on the other hand it only highlights the significance of pure phase correlators
-- which can now be viewed as special functions (building blocks) from which mixed
phase correlators are in turn assembled using only combinatorial quantities:
LR and/or MN coefficients.

Secondly, for the pure phase correlators, we managed to find concise formula
\eqref{eq:pure-phase-final} that finally unifies usual and exotic cases; and makes monomial
matrix models look very similar to celebrated WLZZ models. This opens a way for
various generalizations, such as $(q,t)-$deformation, and Miwa deformation.

\section*{Acknowledgements}
We are grateful to A.Mironov, A.Morozov, N.Tselousov, A.Oreshina and Ya.Drachov
for useful discussions.

A.P. gratefully acknowledges support from the Ministry of Science and Higher Education
of the Russian Federation (agreement no.075-03-2025-662).

\end{document}